\documentstyle[prb,twocolumn,epsf,aps]{revtex} 
\tighten 
\begin{document} 
 
\title{Electron-number statistics and shot-noise suppression 
by Coulomb correlation \\ in nondegenerate ballistic transport} 
\author{O. M. Bulashenko,\cite{byline} 
J. Mateos, D. Pardo, and T. Gonz\'alez} 
\address{ 
Departamento de F\'{\i}sica Aplicada, Universidad de Salamanca, 
Plaza de la Merced s/n, E-37008 Salamanca, Spain} 
\author{L. Reggiani} 
\address{ 
Istituto Nazionale di Fisica della Materia, Dipartimento di Scienza 
dei Materiali, Universit\`a di Lecce \\ Via Arnesano, 73100 Lecce, Italy} 
\author{J. M. Rub\'{\i}} 
\address{ 
Departament de F\'{\i}sica Fonamental, Universitat de Barcelona, 
Av. Diagonal 647, E-08028 Barcelona, Spain} 
\address{\rm (5 May 1997)}
\address{\parbox{14cm}{\rm \mbox{ }\mbox{ }\mbox{ }
Within a Monte Carlo simulation 
we investigate the statistical properties of an electron flow injected 
with a Poissonian distribution and transmitted under ballistic regime 
in the presence of long-range Coulomb interaction. 
Electrons are shown to exhibit a motional squeezing which tends 
to space them more regularly rather than strictly at random,
and evidence a sub-Poissonian statistics with a substantially reduced 
Fano factor $F_n\ll 1$.
The temporal (anti)correlation among carriers 
is demonstrated to be a collective effect which persists over the transit 
of several successive electrons and results in a considerable
(more than one order of magnitude) shot noise suppression.
}}
\address{\mbox{ }}

\maketitle

The study of electronic noise in low-dimensional systems, and mesoscopic 
conductors in particular, has been receiving increasing attention in recent
years. \cite{li90,liu95,birk,reznikov,kumar,jong,{1-3}}
Indeed, by complementing the analysis of current-voltage characteristics, 
the knowledge of statistical properties provides valuable information on 
the microscopic mechanisms ultimately responsible for charge transport.
Uncorrelated carriers exhibiting Poissonian statistics are known to be 
characterized by a shot noise power given by the Schottky formula $S_I=2qI$ 
($q$ being the unit charge and $I$ the average current). 
To a general extent the ideal $2qI$ value can be reduced by  
imparting {\em an anticorrelation} to the electron stream, 
thus making the electron statistics become {\em sub-Poissonian}. 
This can be achieved by exploiting several mechanisms. 
Pauli correlation under degenerate conditions has been investigated in
recent experiments \cite{li90,liu95,birk,reznikov,kumar} and theories 
(see detailed bibliography \cite{jong}). 
A 1/2 noise-suppression factor has been found in a symmetric 
double-barrier junction. \cite{li90,liu95,birk}
For a diffusive conductor a 1/3 suppression has been demonstrated.
\cite{1-3}
The statistics of charge transmitted through a degenerate conductor 
has been found to be binomial, \cite{levitov} while that through a double 
barrier junction is in between a Gaussian and a Poisson distribution. 
\cite{jong96}

Another possibility to incorporate correlations among electrons  
is by means of long-range {\em Coulomb repulsion}. 
The desired effect can be achieved by virtue of a potential barrier 
near an injecting contact, which fluctuates synchronously with 
electron passages through it. 
In this way, an incoming Poissonian flow is converted into an 
outcoming sub-Poissonian flow.
This effect is similar to that leading to shot-noise suppression 
in vacuum diodes and discussed in literature long ago, \cite{ziel54} 
although, in the case of a device, the potential barrier should 
not necessarily be provided by space charge. 
In a semiconductor structure it might be, e.g., a constriction, 
as in the experiment on a quantum point-contact by Reznikov {\em et al.} 
\cite{reznikov} 
We notice that although the importance of Coulomb correlations 
on noise suppression in mesoscopic structures 
has been emphasized, \cite{landauer,buttiker} 
only a few calculations are available, \cite{shot1,shot2} 
and the statistics of charge transmitted
under Coulomb correlations has never been studied to the best of our 
knowledge.

It is the aim of the present paper to address the problem 
of Coulomb correlation by investigating 
the carrier statistics under nondegenerate ballistic transport regime
through a Monte Carlo simulation.
We have found the interesting result 
that just after the potential barrier, self-consistently 
generated by the electron flow, 
the statistical variance of electron-number counts decreases 
with the distance from the barrier. 
Thus, correlated electrons exhibit {\em a motional 
electron-number squeezing},
characterized by a reduced Fano factor $F_n\ll 1$.

For our analysis we consider the following simple model: a lightly doped 
active region of a semiconductor sample sandwiched between two 
heavily doped contacts (of the same semiconductor) subject to an applied 
voltage bias $U$.
The contacts are considered to be ohmic (the voltage drop inside them 
is negligible) and they remain always at thermal equilibrium.
Thus, electrons are injected from the two opposite contacts according 
to a Maxwell-Boltzmann distribution at the lattice temperature $T_L$. 
To exclude correlations due to Fermi statistics, the electron gas 
is assumed to be nondegenerate. 
Electrons are injected with a Poissonian statistics, 
i.e., the time between two consecutive electron injections is 
generated with a probability per unit time
$P(t)=\Gamma_0 e^{-\Gamma_0 t}$, 
where $\Gamma_0=\frac{1}{2} n_c v_{\rm th}S$ is the injection-rate density, 
$n_c$ the carrier concentration at the contact, $S$ the cross-sectional 
area, $v_{\rm th}=\sqrt{2k_B T_L/\pi m}$ the thermal velocity, and $m$ the 
electron effective mass. 
For the simulation we used the following set of parameters:
$T_L$=300 K, $m=0.25 \ m_0$ ($m_0$ being the free electron mass), 
relative dielectric constant $\epsilon$=11.7, active region length
$L=200 \ nm$, and $n_c = 4\times 10^{17} {\rm cm}^{-3}$, 
much higher than the sample doping 
(here taken of $10^{11} {\rm cm}^{-3}$, but the same results are obtained
up to $10^{15} {\rm cm}^{-3}$).
Under these conditions 
the electron dynamics is determined by 
the ratio $\lambda=L/L_{Dc}$, 
where $L_{Dc}$ is the Debye screening length associated 
with the electron injection density. 
The above set of parameters corresponds to $\lambda=30.9$.

We calculate the fluctuating current $I(t)$ and its low-frequency 
fluctuation spectrum $S_I$ by an ensemble Monte Carlo simulator 
self-consistently coupled with a Poisson solver (PS). \cite{shot1}
To illustrate the importance of Coulomb correlation we provide 
the results for two different simulation schemes. 
The first one involves a {\em dynamic} PS: 
the fluctuations of the potential follow self-consistently 
the fluctuations of the space charge by virtue of the Poisson equation. 
The second scheme makes use of a {\em static} PS, which calculates 
the stationary potential profile only, so that carriers move 
in the frozen non-fluctuating electric field. 
Both schemes give exactly the same average current and steady-state 
spatial distributions of all first-order quantities, but they provide 
{\em different noise levels and statistical distributions 
of transmitted carriers.} 

\begin{figure} 
\setlength{\epsfxsize}{8cm} 
\centerline{\mbox{\epsffile{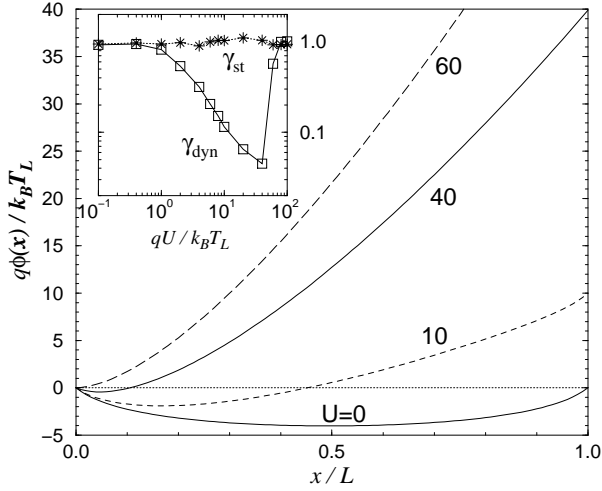}}} 
\caption{Spatial profiles of the normalized potential $\phi(x)$ 
for different applied voltages $U$ (in units of $k_B T_L/q$). 
Inset: shot-noise suppression factor vs voltage $U$ for the dynamic PS 
($\gamma_{\rm dyn}$) and static PS ($\gamma_{\rm st}$), respectively. 
}\label{pot} \end{figure}

Fig.\ \ref{pot} illustrates the potential minimum which, acting as a barrier 
for the electrons moving between the contacts, is shifted 
by changing the bias $U$. 
The corresponding noise-suppression factor $\gamma=S_I/2qI$
is reduced considerably when the dynamic PS is used, 
while for the static PS $\gamma\approx 1$ for all the biases
(see Inset of Fig.\ \ref{pot}). 
The noise suppression becomes particularly pronounced 
at $qU\approx 40 k_B T_L$, when the potential barrier almost vanishes 
completely and its fluctuations affect the transmission 
of the most populated (low-velocity) states of injected carriers. 
At the highest voltages the barrier disappears and no shot-noise 
suppression is observed. 
Note that the shot-noise suppression factor by Coulomb interaction
can achieve as low value as desired in contrast to other mechanisms
(1/3 for diffusive conductors, \cite{1-3} 1/2 for double-barrier junctions 
\cite{jong96}) by increasing the ratio $\lambda=L/L_{Dc}$, i.e., 
by increasing the sample length and/or the carrier concentration 
at the contact, provided the transport remains ballistic.
In the following we fix the bias at $U=40 k_B T_L/q$, 
for which $\gamma\approx 0.045$. 
 
\begin{figure} 
\setlength{\epsfxsize}{8cm} 
\centerline{\mbox{\epsffile{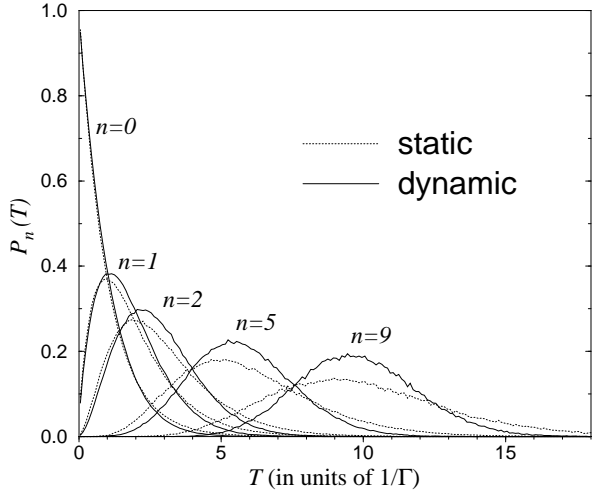}}} 
\caption{Distribution function $P_n(T)$ of electron counting statistics at 
the receiving contact for the static PS (dotted line, Poissonian statistics) 
and for the dynamic PS (solid line, sub-Poissonian statistics). 
}\label{p100}\end{figure}

Having found the level of noise below the ideal Poissonian value, it is 
natural to ask which are the statistical properties of the transmitted 
carriers once they are temporally correlated. 
To address this question we register the times of passage of 
electrons through different sections of the sample at a distance 
$x$ from the injecting contact. 
We register only those going from the cathode to the 
anode, since these are the only electrons contributing to the low-frequency 
current noise. 
With this procedure we are able to study how the carrier 
statistics imposed at $x$=0 (in our case Poissonian) modifies at 
different distances from the cathode. 
For each section $x$ the times of passage of carriers 
$\{t_1,t_2,\dots,t_i,\dots\}$ constitute a stationary stochastic 
point process, which can be described by specifying the random set 
$\{\tau_i\}$, where $\tau_i=t_{i+1}-t_i$ is the time interval between 
events. 
By knowing the set $\{\tau_i\}$, we calculate the distribution function 
$P_n(T)$, which is the probability of detecting $n$ electrons during the 
observation time interval $T$. 
For a Poissonian process all time occurrences are statistically 
independent, which leads to the simple formula 
 
\begin{equation}\label{poiss} 
P_n(T)=\frac{(\Gamma T)^n}{n!}e^{-\Gamma T}, \quad 
n=0,1,2,\dots 
\end{equation} 
where $\Gamma$ is the rate density of events. 
The distribution (\ref{poiss}) is characteristic of uncorrelated 
transport and is tested to correctly describe the carrier statistics at 
the injecting contact at $x$=0. 
The distribution of carriers at the receiving contact depends crucially 
on the PS scheme (see Fig.\ \ref{p100}).
For the static PS $P_n(T)$ is perfectly fitted by the same Poissonian 
formula (\ref{poiss}) as for the injected carriers, 
but with a smaller value of the rate density $\Gamma=\kappa\Gamma_0$, 
with $\kappa\approx 0.65$.
Its reduction is caused by the part of injected 
carriers which is reflected by the potential barrier 
back to the contact. \cite{remark2}
In contrast, for the dynamic PS the distribution function
no longer obeys the formula (\ref{poiss}). 
Fig.\ \ref{p100}
shows the following differences between the dynamic and static PS: 
(i) for each $n\geq 1$ the maximum of the distribution is shifted to 
longer $T$; (ii) the probability distribution is narrowed; 
(iii) the higher the index $n$, the more the dynamic case deviates 
from the static one, the distribution profiles becoming 
more symmetrical and closer to a Gaussian shape. 
We remark that the difference at point (ii) can be interpreted as a {\it 
motional squeezing of electron number} 
and corresponds to a higher regularization of the carrier passage due to 
correlation among them. 
 
\begin{figure} 
\setlength{\epsfxsize}{7cm} 
\centerline{\mbox{\epsffile{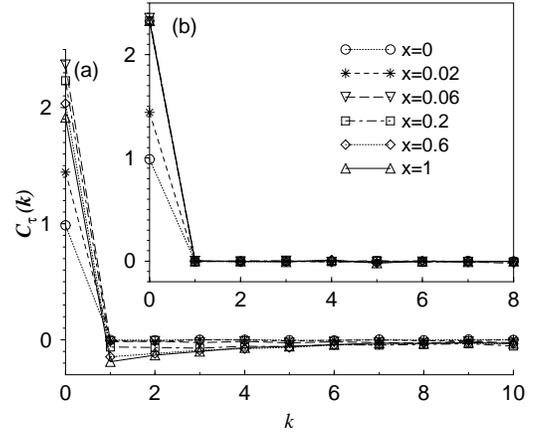}}} 
\caption{Correlation function of inter-event times as a function of 
the number of successive events at different distances $x$ 
from the injecting contact 
for the dynamic (a) and static (b) PS, respectively. 
For a better view discrete points are connected by lines. 
$x$ is in units of the sample length $L$. 
}\label{cor} \end{figure}

\begin{figure} 
\setlength{\epsfxsize}{7cm} 
\centerline{\mbox{\epsffile{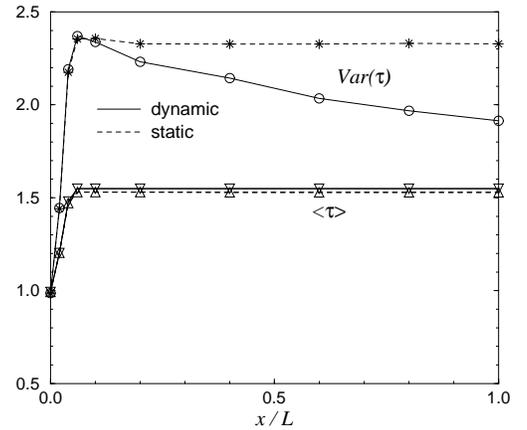}}} 
\caption{Mean and variance of inter-event times as a function of 
distance $x$ from the injecting contact.
$\tau$ is multiplied by the injection rate $\Gamma_0$.
}\label{mom} \end{figure}

The temporal correlation among carriers can be characterized by 
the autocorrelation function of the inter-event times as a function of 
the number of successive events defined as 
 
\begin{equation}\label{autocor} 
C_{\tau}(k)=\langle 
(\tau_j-\overline{\tau})(\tau_{j+k}-\overline{\tau})\rangle, \quad 
k=0,1,\dots 
\end{equation} 
Here $\overline{\tau}$ is the mean of the random variable $\tau_i$, and 
angular brackets denote average over the reference index $j$. 
Fig.\ \ref{cor} shows the function $C_{\tau}(k)$ calculated for the static 
and dynamic PS using the random sets $\{\tau_i\}$ registered at 
different positions $x$ along the sample. 
Firstly we notice that for the static PS the correlation 
function is zero for all $k\neq 0$ [Fig.\ \ref{cor}(b)]. 
This means that the absence of correlation among carriers imposed by 
the Poissonian injection is preserved at any section of the sample. 
Indeed, being the electric potential frozen no additional correlation is 
introduced inside the system. 
The only nonzero value is the variance 
$C_{\tau}(0)={\rm Var}(\tau)=\overline{\tau^2}-\overline{\tau}^2$, which 
increases with $x$ until $x$ coincides with the minimum of the potential 
profile (here the carrier flow is 
decreased due to the reflection), and then saturates for larger values of 
$x$ (see Fig.\ \ref{mom}). 
In contrast, for the dynamic PS $C_{\tau}(k)$ is in general 
nonzero for $k\neq 0$ and, moreover, it is {\em negative} by decaying 
to zero at increasing values of the index $k$ [Fig.\ \ref{cor}(a)]. 
Negative values of $C_{\tau}(k)$ start to appear just after the 
potential minimum, their magnitude tending to increase for longer distances. 
The negative sign means that once a time interval $\tau_m$ 
longer than the average value $\overline{\tau}$ is registered, 
to compensate such a deviation the successive $\tau_{m+1}$ is expected 
to be shorter than $\overline{\tau}$. 
The mechanism responsible for this compensating behavior is the 
self-consistent redistribution of the potential which, by varying the barrier 
height introduces a correlation into the carrier flow. 
{}From Fig.\ \ref{cor}(a) we estimate that the number of consecutive 
carrier passages over which the correlation persists is $\approx 6$ 
[i.e., $C_{\tau}(k)\approx 0$, for $k>6$]. 
It should be stressed that the mean time between consecutive electrons 
is not influenced by the long-range Coulomb correlation, 
since $\overline{\tau}$ is related to the average current. 
Fig.\ \ref{mom} shows that for both the static and dynamic PS 
$\overline{\tau}$ increases in the region before the barrier 
($x\leq 0.06 L$), due to the loss of the part of injected electrons 
which are reflected by the potential barrier, and then saturates. 
The Coulomb correlation is mirrored by the second moment ${\rm 
Var}(\tau)$ which, for the dynamic PS, is found to decrease with 
distance just after the barrier (see Fig.\ \ref{mom}). 
 
\begin{figure} 
\setlength{\epsfxsize}{7cm} 
\centerline{\mbox{\epsffile{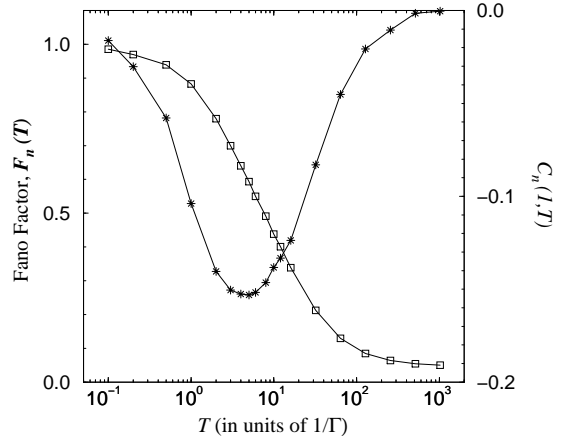}}} 
\caption{ 
Fano factor $F_n(T)$ (left axis, squares), 
and correlation function of the number of counts between two 
consecutive observation intervals, $C_n(1,T)$, normalized by 
$\overline{n}^T$ (right axis, stars), 
calculated at the output contact within the dynamic scheme. 
}\label{min} \end{figure}

The autocorrelation function $C_{\tau}(k)$ refers to the statistical 
distribution of inter-event times $\{\tau_i\}$ which, in turn, is 
related to the operation mode of registering the carrier arrival times. 
Another operation mode is the number-counting, 
for which the number of carriers $n_i^T$ in a given time 
interval $T$ is registered. 
For this mode the statistical information is carried by the random variable 
$\{n_i^T\}$ which depends on the observation time $T$ as a parameter. 
The first-order statistics of the random set $\{n_i^T\}$ is described by 
the distribution function $P_n(T)$ shown, for example, in Fig.\ \ref{p100} 
for the carriers at the receiving contact. 
A better insight into the carrier correlation can be obtained 
by calculating the correlation function 

\begin{equation}\label{corn} 
C_n(k,T)=\langle 
(n_j^T-\overline{n}^T)(n_{j+k}^T-\overline{n}^T)\rangle, \quad k=0,1,\dots, 
\end{equation} 
representing the second-order statistics. 
Here, $\overline{n}^T$ is the mean number of counts during 
the observation time interval $T$. 
For fixed $T\sim\overline{\tau}$, $C_n(k,T)$ behaves like $C_{\tau}(k)$, 
i.e., it is negative for $k\geq 1$ with its lowest value at $k$=1 
[like in Fig.\ \ref{cor}(a)]. 
Therefore, $C_n(1,T)$ can be a good measure of the correlation 
between the number of counts. 
This quantity, normalized to the mean number of counts $\overline{n}^T$, is 
shown in Fig.\ \ref{min} for the statistics registered at the receiving 
contact under dynamic regime.  
$C_n(1,T)$ is seen to display a minimum as a function of $T$ 
(indicating the maximum of correlation) for $T\approx 5/\Gamma$. 
This means that the maximum of the correlation occurs when the observation 
time interval $T$ is chosen such that the number of counts is 
$\overline{n}^T\approx 5$. 
In fact, this is another estimation of the number of consecutive carriers 
over which the correlation persists, in addition to that obtained from the 
approximate decaying range of $C_{\tau}(k)$ of Fig.\ \ref{cor}(a) 
($k\approx 6$). 
 
The correlation function for $k$=0 gives the variance of the number of 
counts $C_n(0,T)={\rm Var}(n^T)=\overline{(n^T)^2}-(\overline{n}^T)^2$. 
In Fig.\ \ref{min} we present also the ratio of the variance to the mean 
$F_n(T)={\rm Var}(n^T)/\overline{n}^T$ known as the Fano factor in photon 
counting statistics. \cite{saleh92}
It characterizes the degree of deviation of the counting 
statistics from the Poissonian one, whenever $F_n(T)\neq 1$.
We have found that for the static PS $F_n(T)=1$ always, while for the 
dynamic PS it decreases with $T$ (see Fig.\ \ref{min}). 
Notice that if the observation time $T\ll 1/\Gamma$, 
$F_n(T)\approx 1$, i.e. Coulomb correlation among carriers cannot be detected 
by any means, and the observed statistics of carriers is always Poissonian. 
For $T\gg 1/\Gamma$ the correlation between the counting intervals is lost 
[$C_n(1,T)\to 0$], while the Fano factor approaches 
its minimum value at the shot-noise suppression factor
$F_n(\infty)\equiv\gamma\approx 0.045$ (Fig.\ \ref{min}).

In conclusion, within a Monte Carlo scheme we have investigated the 
statistics of carriers which are Poissonian injected and 
transmitted through a nondegenerate ballistic structure in the presence of 
long-range Coulomb interaction. 
The sub-Poissonian statistics exhibited by carrier number fluctuations 
registered at the receiving contact provides an interesting 
example of motional squeezing in analogy with photon satististics. 
The negative value of the correlation function of both: 
time intervals between consecutive 
electrons passage, and number of counts in successive observation time 
intervals, mirrors the temporal correlation of transmitted carriers 
introduced by the presence of space charge. 
The small value of the Fano factor ($F_n\ll 1$) here found, besides being of
fundamental interest, offers interesting perspectives in communication 
systems.
Indeed, a sub-Poissonian electron flux can generate a 
photon-number-squeezed light which, in turn, can be used to enhance a 
channel capacity to transmit information in optical communication systems.
\cite{saleh92}
 
This work has been supported by the Comi\-si\'on Interministerial de 
Ciencia y Tecnolog\'{\i}a through the project TIC95-0652, 
by the DGICYT of the Spanish Government under grant PB95-0881,
and Ministero dell' Universit\`a e della Ricerca Scientifica 
e Tecnologica (MURST).

\begin{references}
\bibitem{remark2} 
In fact, the current for the present applied voltage is $I=\kappa I_s$, 
with $I_s=q\Gamma_0$ the saturation current.
\end{references}
\end{document}